\documentclass[12pt,preprint]{aastex}

\begin{document}

\title{Wandering Stars: an Origin of Escaped Populations}

\author{Maureen Teyssier\altaffilmark{1},
Kathryn V. Johnston\altaffilmark{1}, Michael M. Shara\altaffilmark{2}}
\altaffiltext{1}{Department of Astronomy, Columbia University, Pupin Physics Laboratory, 550 West 120th Street, New York, NY 10027, USA}
\altaffiltext{2}{American Museum of Natural History, 79th Street and Central Park West, New York, NY 10024-5192, USA}

\begin{abstract}
We demonstrate that stars beyond the virial radii of galaxies may be generated by the gravitational impulse received by a satellite as it passes through the pericenter of its orbit around its parent.  These stars may become energetically unbound (escaped stars), or may travel to further than a few virial radii for longer than a few Gyr, but still remain energetically bound to the system (wandering stars).  Larger satellites (10-100\% the mass of the parent), and satellites on more radial orbits are responsible for the majority of this ejected population. Wandering stars could be observable on Mpc scales via classical novae, and on 100 Mpc scales via SNIa.  The existence of such stars would imply a corresponding population of barely-bound, old, high velocity stars orbiting the Milky Way, generated by the same physical mechanism during the Galaxy's formation epoch.  Sizes and properties of these combined populations should place some constraints on the orbits and masses of the progenitor objects from which they came, providing insight into the merging histories of galaxies in general and the Milky Way in particular.
\end{abstract}


\section{Introduction}
One of the triumphs of the last decade is orders of magnitude increase in the numbers of stars catalogued by large sky surveys such as the Sloan Digital Sky Survey and the Two Micron All Sky Survey.  
These catalogs contain significant samples of stars out to $\sim$100kpc from the center of the Galaxy, allowing these regions to be probed through number counts to far lower surface brightness than previously possible and resulting in the discovery of numerous new dwarf satellite galaxies \citep[e.g.][]{2005ApJ...626L..85W,2006ApJ...647L.111B}, as well as tidal tails from dwarf galaxies possibly disrupted long ago \citep[e.g.][]{2003ApJ...599.1082M, 2005MNRAS.357...17B}. 

We move from this decade of large sky surveys into a decade of continual all-sky surveying with the introduction of the Panoramic Survey Telescope \& Rapid Response System \citep[Pan-STARRS, see][]{2004SPIE.5489...11K} and the prospect of the Large Synoptic Survey Telescope \citep[LSST, see ][]{ 2007AAS...210.6605I}. These surveys will be sensitive to ever fainter magnitude objects and hence probe ever lower surface brightness structures in ever increasing volumes of space.  The repeated nature of these surveys also adds the extra dimension of time, enabling large-sample statistical studies of variable phenomena in a way that was not previously possible.  For example, \cite{2006AJ....131.2980S} points out that LSST will detect many new classical novae (hereafter CN).  While interesting in their own right, the special relationship between the peak brightness and decline time relationship of CN, coupled with their high luminosity, will allow us to map a volume out to 40 Mpc from Earth. 
This will be the first map of stars between galaxies and outside of clusters or groups of galaxies (i.e. intercluster stars) ever produced.  But should we expect to find anything there?

Red giant stars, planetary nebulae and classical novae  have been detected outside of galaxies but {\it within clusters} \citep[i.e. intracluster light - hereafter ICL, for example see][]{2004MNRAS.355..159W,2005ApJ...618..692N}.
These populations are thought to begin their lives in galaxies: 
stars are liberated from their original hosts during galactic collisions, galactic harassment and tidal shredding, which generate tails of debris that can be ripped from the galaxy system by the tidal field of the cluster \citep{1996Natur.379..613M}.   
This evolutionary picture is confirmed by the general agreement of observations and simulations on the level of ICL \citep{2004ApJ...607L..83M},which photometric studies define as $0  -  50\%$ of the total cluster light \citep{2004bdmh.confE..26A,2004IAUS..217...86F,2004MNRAS.355..159W}.   
The same dynamical processes can be appealed to on smaller scales  to explain the existence of ICL around  less massive, looser clusters and galaxy groups \citep{2004bdmh.confE..26A}, as well as diffuse stellar halos around individual galaxies \citep{2005ApJ...635..931B}.
Thus stellar halos and the ICL can be thought of  as testaments to galactic interactions during the epochs of galaxy and cluster formation.  

We explore whether dynamical interactions could plausibly lead to stars {\it beyond} the virial radii of the parent galaxy, group or cluster, and, if so, what the existence of such a population might tell us 
about the nature of hierarchical structure formation.
\citet{1992ApJ...397L..75S} have already shown that N-body simulations of mergers of massive galaxies can give rise to unbound particles that form a gaussian tail to the energy distribution, in agreement with a statistical mechanical description of violent relaxation.
Here, we develop a simple description of interactions that allows us to illustrate how the size of the escaped population should depend on the mass ratio and orbits of the progenitor systems. 
In \S \ref{Mechanism for Escape} we establish that a tidal impulse at perigalacticon is a viable path for  stars from satellite dwarf galaxies to escape from the combined dark matter potentials of the parent and satellite galaxies.  
 In \S \ref{Illustration} we first use restricted three-body simulations to test the mechanism described in \S \ref{Mechanism for Escape} and then analyze N-body results to establish trends, and quantitative expectations for the escaping population.
 In \S \ref{Discussion} we discuss the implications of our results for future surveys of giant stars, CN, supernovae and high velocity stars.  
 In \S \ref{Conclusions} we summarize our conclusions.

\section{Mechanism for Escape} \label{Mechanism for Escape}

Consider a satellite orbiting in a parent potential $\Phi(r)$, at radius $r$ from the center of the parent, moving with velocity $\mathbf{v}$, along an orbit of energy $E = \frac{1}{2}\mathbf{v}\cdot \mathbf{v} + \Phi(r)$. 
As the satellite orbits, a star at position $\mathbf{x}$ relative the center  will experience a slightly different acceleration due to the parent than the satellite as a whole. In the impulsive regime, the resultant relative velocity change between the star and the satellite can be approximated as $\Delta {\mathbf v} \sim \Delta {\mathbf a}_{\rm peri} t_{\rm enc}$ where $t_{\rm enc}=r_{\rm peri}/v_{\rm peri}$ is the duration of an encounter of pericentric distance $r_{\rm peri}$ and speed $v_{\rm peri}$, and $\Delta \mathbf{a}_{\rm peri}$ is the difference in acceleration experienced by the satellite and star at pericenter. 
Expanding $\mathbf{\Delta v} $ in powers of $\mathbf x$ we have:
\begin{equation}
\mathbf{\Delta v} \sim  {\mathbf a}_{\rm peri} t_{\rm enc} = (\mathbf{x} \cdot \mathbf{\nabla}) \nabla \Phi(r)t_{\rm enc} \;
\sim -GM_{\rm peri}\left(\frac{\mathbf{x}}{|\mathbf{r_{\rm peri}}|^3}-3\frac{\mathbf{r_{\rm peri}}(\mathbf{x} \cdot \mathbf{r_{\rm peri}})}{|\mathbf{r_{\rm peri}}|^5}\right)t_{\rm enc}
\label{diffaccel}
\end{equation}
\noindent where $M_{\rm peri}$ is the enclosed mass of the parent at pericenter.  The first and second terms in equation (\ref{diffaccel}) are responsible for stretching along $\mathbf{r}$ and compression perpendicular to $\mathbf{r}$, respectively.
 
This impulsive velocity change corresponds to a change in orbital energy
\begin{equation}  \Delta E =   \mathbf{v}\cdot \mathbf{\Delta v} +  \frac{(\Delta v)^2}{2}
\label{all}
\end{equation}
When summed over all stars in the entire satellite, the first term in equation (\ref{all}) typically vanishes due to symmetry, and the last term represents the dynamical heating.  In contrast, for an individual star, the last term is usually negligible compared to the first.  The maximum increase in orbital energy will be for stars barely bound to (i.e. at large $r_{\mathbf *}=|{\mathbf x}|$) and trailing the satellite along its orbit (i.e. where the direction of the impulse $\Delta {\mathbf v}$ is aligned with the orbital motion), where 
\begin{equation}
\Delta v \sim \frac{2GM_{\rm peri}}{r_{\rm peri}^3}  r_{\mathbf *} t_{\rm enc} 
\sim \frac{2GM_{\rm peri}}{r_{\rm peri}^2}\frac{r_{\mathbf *}}{v_{\rm peri}}.
\label{simplediff}
\end{equation}
Such a trailing star could move to an an orbit that escapes entirely from the combined parent-satellite system if $\Delta E \sim v_{\rm peri} \Delta v > |E|$ or 
\begin{equation}
\Delta v > \frac{|E|}{v_{\rm peri}}.
\label{einequal}
\end{equation} 
In combination, equation (\ref{einequal}) and equation (\ref{simplediff}) yield a constraint on the
minimum extent of stars in the satellite system for an escaping population to be produced:
$r_{\mathbf *}>r_{{\mathbf *},\rm min}$ where
\begin{equation}
r_{\mathbf{*}, \rm min}= \frac{|E|}{2GM_{\rm peri}/r_{\rm peri}} \;  r_{\rm peri}.
\label{rcondition}
\end{equation}
  
\begin{figure}[htbp]
\begin{center}
\includegraphics[width=0.5\textwidth]{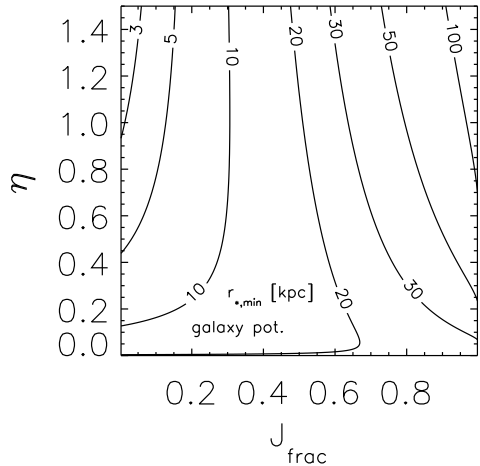}
\caption{Estimated minimum light radius [kpc] of the satellite necessary for unbinding tracer particles from perigalacticon of a Milky-Way-size haloas a function of binding energy (parameterized by $\eta$ --- see text) and orbital eccentricity ($J_{\rm frac}=0$ for radial orbits and $J_{\rm frac}=1$ for circular orbits).
}
\label{allcalc}
\end{center}
\end{figure}

We examine whether equation (\ref{einequal}) will be satisfied for any realistic galactic encounters by estimating $r_{{\mathbf *}, \rm min}$ for orbits in a Milky Way-like parent halo that follows a Navarro-Frenk-White potential of virial mass $m_{\rm vir}=1.4 \times 10^{12} M_{\odot}$, virial radius of $r_{\rm vir}=$273 kpc\, and characteristic radius $r_c$=20.7 kpc.  The functional form of the NFW is:
\begin{equation}
m(r) = m_{\rm c}\left( \log\left( \frac{r +r_{\rm c}} {r_{\rm c}}\right) - \frac{r}{r + r_{\rm c}}\right)
\label{mass}
\end{equation}
where the mass within distance $r$ is $m(r)$ and $m_{\rm c} = 4 \pi \delta_{c} \rho_{c} {r_c}^3$ is the mass within the characteristic radius  \citep{1994MNRAS.267L...1N}. 
Fig.\ref{allcalc} shows numerically solved contours of $r_{\mathbf{*}, \rm min}$ as a function of circularity, $J_{\rm frac} = \frac{J}{J_{\rm circ}}$, and fractional binding energy parameter, $\eta\frac{R_{\rm circ}}{r_{\rm vir}}$, for a satellite orbit of energy $E$ and angular momentum $J$.   $J_{\rm circ}$ and $R_{\rm circ}$ are the angular momentum and radius of a circular orbit of the same energy $E$.  Satellite orbits become more loosely bound with increasing $\eta$ and more eccentric with decreasing $J_{\rm frac}$.
 From Fig.\ref{allcalc}, $r_{{\bf *}, \rm min}$ depends weakly on binding energy, but strongly on orbital eccentricity.  For any $\eta$ or $J_{\rm frac}$, more massive (i.e. of greater stellar extent) satellites are likely to create a greater number of E/W stars than a smaller satellite.  A satellite on an eccentric orbit will contribute more stars than the same satellite on a less eccentric orbit. The lowest contours on Fig.\ref{allcalc} overlap with the typical extent of stars in Local Group dwarf galaxies.  As we observe satellites which meet criteria in Fig.\ref{allcalc} today, logically, this is indeed a plausible mechanism for ejecting stars from a Milky-Way-sized halo during its formation epoch. 

Analogous calculations performed for an NFW potential on cluster scales, with $M_{\rm vir}=1 \times 10^{14} M_{\odot}$, show contours of similar shape but more than a factor 10 increase in amplitude.  While these scales are much greater than the sizes of typical galactic disks, these results do allow that the same mechanism may liberate the most loosely bound parts of galaxies from the cluster --- the stellar halos of spirals and the outer edges of ellipticals.

\section{Illustrations of Escape} \label{Illustration}

\subsection{Restricted 3-Body Simulations} \label{3body}

A restricted 3-body code follows the orbits of stars as a dark matter satellite orbits its parent galaxy. Stars may (``escaped'' stars --- E) or may not (``wandering'' stars --- W) become energetically unbound from the parent-satellite system, but the two classes are likely to be observationally indistinguishable; a wandering star can take a jaunt beyond the virial radius that lasts for a significant fraction of the age of the universe. Note- the intention in this section is not to estimate the size of the E/W population, but to confirm the predicted escape mechanism (\S\ref{Mechanism for Escape}).

Our restricted 3-body model consists of two orbiting, analytical, nonevolving NFW dark matter halos, and 10,000 massless ``tracer'' particles which represent stars around the smaller of the two halos (the ``satellite''). Simulations are performed in two pieces --- orbits for interacting NFW halos are calculated first, then tracer particles are run in the changing potential. The code uses leapfrog integration. Orbital precision is ensured by comparing results for the same parameter sets calculated with orders of magnitude change in time resolution.

The Milky Way-like (as described in \S\ref{Mechanism for Escape}) parent NFW is identical in all runs.  Simulations run with satellites of various characteristic masses ($m_c$, of 10\%, 1\%, and .1\% of the parent) and scale radii  \citep[$r_c$, of 7 kpc, 3 kpc, and 1 kpc, drawn from distributions seen at $z = 0$ in cosmological simulations ---][]{2001MNRAS.321..559B} are placed on orbits with various $R_{circ}$ (130 kpc, 200 kpc) and $\eta$ (.05, 0.1, 0.5).  Parent and satellite dark matter halos are truncated at their respective tidal radii at initial apogalacticon.

Since we are interested in generating escaping stars, massless particles are not chosen to reproduce a fixed density distribution, but rather to explore regions of phase-space that are most likely to be stripped. They are placed randomly with a spatially uniform distribution around the satellite, out to its tidal radius.  Their velocities are initially spatially isotropic, with amplitudes forming an uniform energy distribution up to the escape energy of the satellite at a given radius $E_{\rm esc,sat}(r)$, i.e. the vast majority of stars should escape from the satellite.  We eliminate particles unbound within the first 1/10 of the orbit; they are uninteresting for analysis of evolution near pericenter.

\begin{figure}[h]
\begin{center}
\includegraphics[width=.9\textwidth]{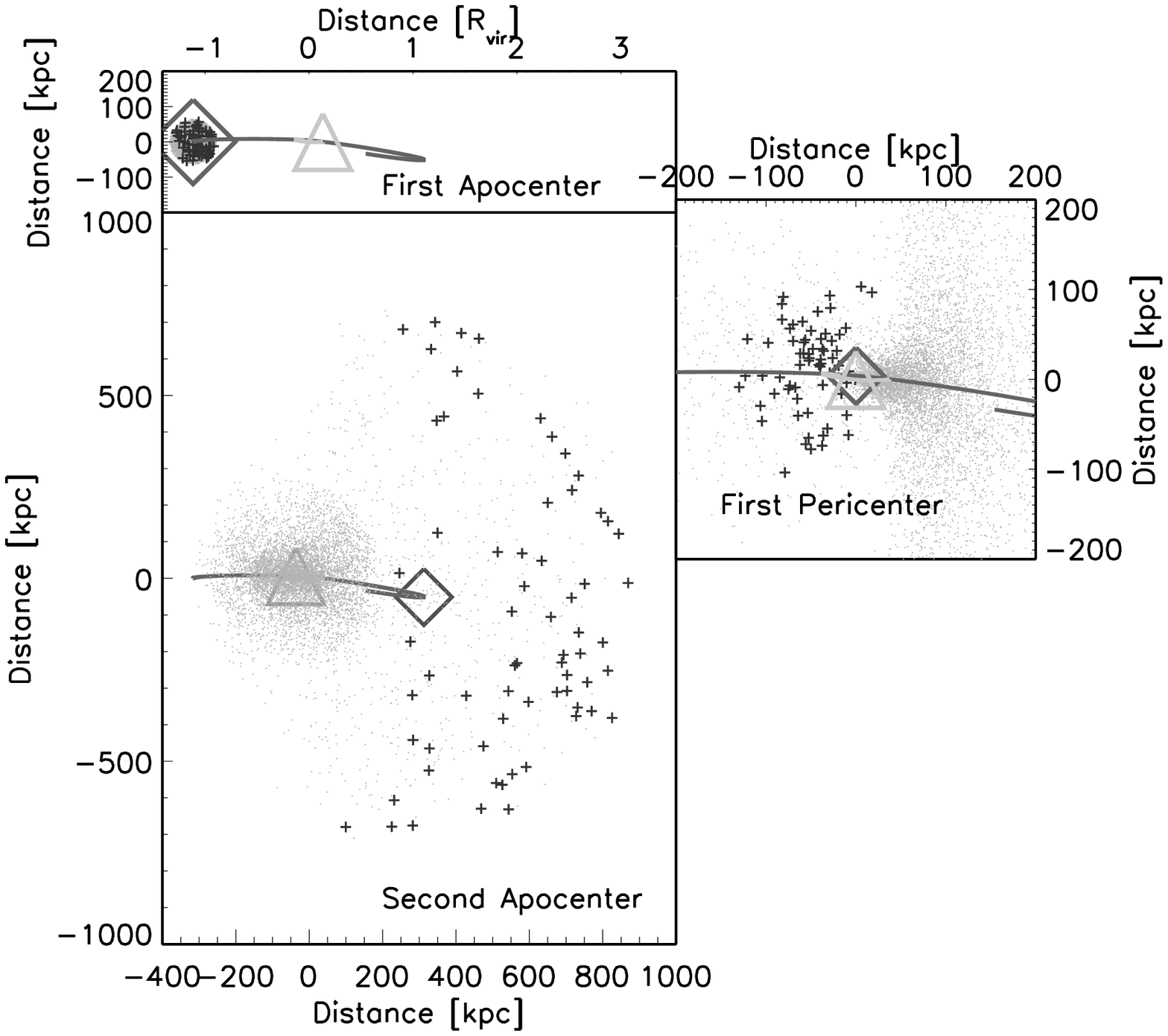}
\caption{
Projection of particle positions in a 3-body simulation onto the orbital plane. Future E/W particles are marked with black stars. The center of the satellite is a diamond, its orbit is dark grey. The parent NFW is represented by a light grey triangle with its orbit shown in light grey.}
\label{xyplane}
\end{center}
\end{figure}

Figure \ref{xyplane} is an example satellite from these runs, with $m_c$ 10\% that of the parent, on an orbit with $R_{circ}$ = 200 kpc and $\eta$ = 0.05.  
The satellite is mostly stripped of it's initially uniform cloud of test particles (gray) by pericenter. Test particles that are ahead of the satellite along its orbit at the first pericenter form a cloud around the center of the parent potential by second apocenter. In contrast,  black crosses highlight the positions of the subset of particles that form the E/W population --- defined as those that are ejected and remain outside the virial radius of the system for more than 2 Gyears. While this subset follows the same isotropic distribution as the non-escapers at apocenter, they clearly lag the satellite along its orbit at pericenter. Every run which generates wandering stars demonstrates this behavior, confirming the dynamical picture proposed in \S \ref{Mechanism for Escape}; the E/W population can be ejected by an implusive velocity change at pericenter aligned with the orbital motion.

\subsection{N-Body Simulation}\label{BJ05}

While the restricted 3-body models successfully illustrate the mechanism for generating E/W stars, they are of limited use in assessing the size of this population or trends in progenitor properties for four reasons:  satellites do not self-consistently lose mass as they orbit; star particles are not embedded deep within their host dark matter halos; parent potentials do not grow; and satellite's orbits do not evolve due to dynamical friction with the parent. (Note- none of these effects would invalidate our description of the formation mechanism.) 

We turn our attention to cosmological N-body simulations \citep{2005ApJ...635..931B} of satellite accretion, where mass loss from the satellite is modelled self-consistently, stars are distributed more realistically, an estimate of dynamical friction is explicitly included, and parent potentials grow analytically. The original aim of these simulations was to examine the formation of the stellar halo of the Milky Way. The set includes 11 stellar halo models, each formed from the superposition of simulations of individual satellite accretions, whose masses, accretion times and orbits were chosen at random from cosmologically motivated distributions. Gravitational influence of the parent galaxy is calculated from analytic functions which represent disk, bulge and dark matter halo components and that grow smoothly over time (tied to a randomly generated accretion history) to form a Milky Way type galaxy today. Dark matter in each satellite is represented by $10^5$ equal-mass particles, with positions and velocities initially chosen to follow an equilibrium NFW distribution. Masses and scales of these NFW halos are drawn from those seen in fully self-consistent cosmological simulations of structure formation. Stars are ``painted-on'' by assigning a variable mass-to-light ratio to each dark matter particle in such a way that they follow an equilibrium King model, whose total luminosity, spatial and velocity scales are normalized to match observed scaling relations of Local Group dwarf galaxies.

Our 11 stellar halo models all contain an E/W population with an average of 5.5\% (and ranging between 2.5 and 10\%) of the stars beyond the virial radius of the host today. Since these models only follow the stellar halo component of the Galaxy ($\sim 10^9 M_\odot$ in stars), this represents a much smaller fraction of the total light of the Galaxy --- of order 0.05\%. This is a lower limit to the E/W population since it does not include those that might be contributed from the parent galaxy, or from disk components of satellites during more major mergers. 

\begin{figure}[h]
\begin{center}
\includegraphics[width=0.4\textwidth]{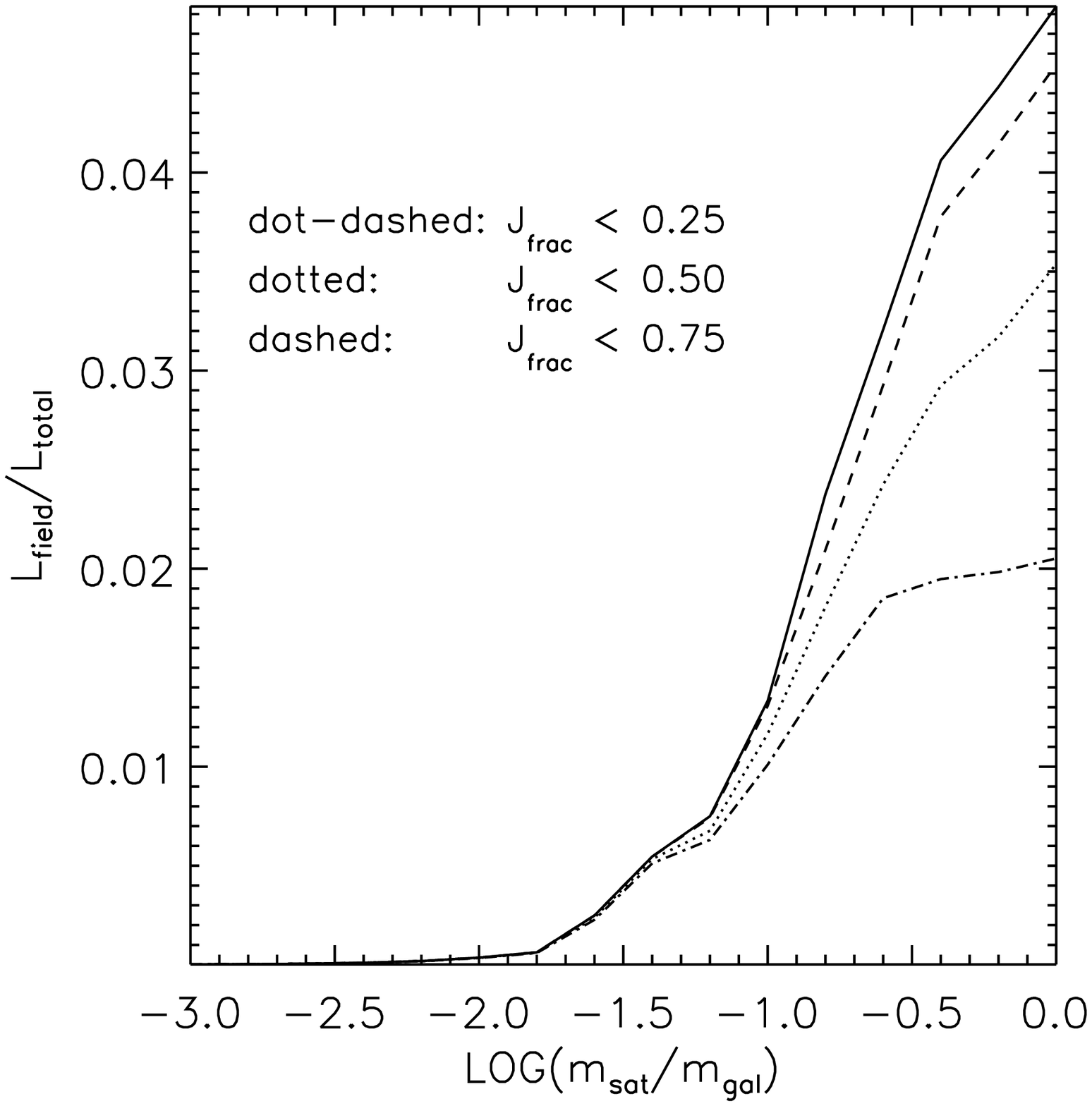}
\caption{Solid line shows the cumulative fraction of the E/W population (i.e. number of E/W divided by the total number of stars falling in to the parent) contributed by satellites less than a given mass.  The dot-dashed line shows the most eccentric orbits.  Dotted and dashed lines shows the additional contribution from satellites on orbits that are increasingly circular.}
\label{fractions}
\end{center}
\end{figure}

We ask what type of interactions are responsible for the E/W population in these simulations. In Figure \ref{fractions} we look at the fraction of stars defined as E/W, averaged over all 11 stellar halo models as a function of satellite mass and orbital properties.
The solid line plots the cumulative fraction of E/W stars (i.e. number of E/W divided by the total number of stars falling in to the parent) from all satellites less than a given mass. The dot-dashed/dotted/dashed lines are the subset of satellites with $J_{\rm frac} < 0.25/0.5/0.75$ --- the orbit distribution for these satellites was originally nearly uniform in $J_{\rm frac}$. This figure confirms the trends anticipated in \S \ref{Mechanism for Escape}; the bulk of the E/W population is produced by: (i) larger mass satellites since they are intrinsically more extended and more stars satisfy the criterion $r_{\bf *}> r_{\mathbf{*}, min}$; and (ii) satellites on more eccentric orbits with smaller pericenters that minimize $ r_{\mathbf{*}, min}$. 
In particular, by comparing the solid and dot-dashed lines we see that for all masses, much more than 50\% of the population is produced from the 50\% of objects on the most eccentric orbits. Also, at the very lowest masses the dot-dashed line converges with the total: smaller satellites need to be on increasingly eccentric orbits in order to contribute to the E/W population.

\section{Discussion}\label{Discussion}

Our results indicate the plausible existence of a population of isolated stars thrown out beyond the virial radii of Milky-Way-size dark matter halos. We expect that most of this population was generated during the epoch of structure formation on various scales. While these results should be confirmed by simulations of structures forming in a fully self-consistent cosmological context, our estimates suggest a lower limit of 0.05\% of stars could be in this population of E/W stars, and this number could be revised significantly upwards once the contribution from the disk component in equal-mass mergers is included.

There are promising avenues for detecting such a population.  \cite{2006AJ....131.2980S} has pointed out that near-future all-sky, repeated surveys should be able to map the distribution of CN out to the Virgo cluster. Assuming an E/W population of $\sim$10\% (from the levels of ICL) he calculates a detection rate of 100's to thousands during the first few years of LSST for a typical stellar population.  Such surveys would also be sensitive to SNIa, an even brighter, though rarer,  variable phenomena: extragalactic SNIa  \citep{2009CBET.1682...13G,1981AJ.....86..998S,2003AJ....125.1087G} may not have been truly represented in more targeted observations.  

While our discussion has focussed on stars, numerical simulations of structure formation on both cluster and individual galaxy scales have also noted that satellite galaxies can sometimes be accreted onto parent potentials in groups \citep{2006ApJ...648..936R,2008MNRAS.385.1365L}. A low mass satellite, initially only loosely bound to an infalling group could potentially be ejected by the same mechanism described in \S \ref{Mechanism for Escape}. Indeed, populations of galaxies at several virial radii away from their prior parent \citep[dubbed backsplash galaxies by][]{2005MNRAS.356.1327G} have been seen in many simulations \citep{2005MNRAS.356.1327G,2009ApJ...692..931L}. Moreover, there are now dwarf galaxies observed within the Local Group to be moving at close to or even above escape speed \citep{2007ApJ...670L...9M,2007ApJ...662L..79C}.

Lastly, an interesting implication of our results is that, accompanying our E/W population, there must be a population of stars that are kicked to barely bound orbits. At their apocenters, these would contribute to our ``wandering'' population, but as they oscillated back through the Milky Way they  could be detectable as {\it old} high-velocity stars (HVS) in the stellar halo, left over from the epoch of galaxy formation. Current surveys have  concentrated on mapping the young HVS population, in part because they are easy to select photometrically \citep[i.e. from unusual blue colors and faint magnitudes at high Galactic latitude, see][]{2006ApJ...647..303B}. This young population is thought to originate from multiple-star encounters with the Galactic Center black hole --- though \citet{2009ApJ...691L..63A} have shown using N-body simulations that such speeds can be produced in satellite/parent galaxy encounters, presumably by the mechanism we describe analytically here. Systematic surveys of the stellar halo velocity distribution of the older (redder) population are just now producing results \citep{2009ApJ...697.1543K} and could provide an interesting additional limit on the E/W population.

\section{Conclusions}\label{Conclusions}

We've outlined how tidal interactions can eject stars which are loosely bound to a satellite galaxy to beyond the virial radius of the satellite's parent.  We determined that more massive satellites contribute the most to this escaped (entirely unbound from the satellite/parent system) or wandering (bound, but traveling beyond 2 virial radii for more than $\sim$Gyr) population of stars, as do satellites on highly radial orbits.  The majority of this population of stars originated during the epoch of galaxy formation, when large interacting satellites were more frequently infalling on radial orbits, approximately from a redshift of around 4 to 3.  If this is indeed the case, we expect an old, uniform (to first order) population of escaped and wandering stars to exist beyond a few virial radii of the Milky Way.  Subcategories (e.g. classical novae, supernovae) of this population should be detectable by LSST and Pan-STARRS. We estimate a lower limit of 0.05\% of stars to be members of this population. An accurate number prediction warrants further investigation using full cosmological simulations which follow interacting subhalos as they form the parent halo. 

"Not all those that wander are lost"-Tolkien

\acknowledgements
Thanks to James Bullock for contributions to the simulations analyzed in this work,  and Josh Barnes for inspiration about equal mass mergers.  Work was supported in part by NSF grant AST-0734864.

\end{document}